\documentclass[
preprint,longbibliography,
superscriptaddress,
 amsmath,amssymb,
prappl,
]{revtex4-2}

\newcommand\beq{\begin{equation}}
\newcommand\eeq{\end{equation}}

\usepackage{graphicx}
\usepackage{dcolumn}
\usepackage{bm}
\usepackage{xcolor}
\usepackage{soul}

\begin{document}


\title{Exceptional Points in Flat Optics: A Non-Hermitian Line-Wave Scenario}

\author{Massimo Moccia}
\affiliation{Fields \& Waves Lab, Department of Engineering, University of Sannio, I-82100 Benevento, Italy}
\author{Giuseppe Castaldi}
\affiliation{Fields \& Waves Lab, Department of Engineering, University of Sannio, I-82100 Benevento, Italy}
\author{Francesco Monticone}
\email{francesco.monticone@cornell.edu}
\affiliation{School of Electrical and Computer Engineering, Cornell University, Ithaca, New York 14853, USA}
\author{Vincenzo Galdi}
\email{vgaldi@unisannio.it}
\affiliation{Fields \& Waves Lab, Department of Engineering, University of Sannio, I-82100 Benevento, Italy}

\date{\today}


\begin{abstract}
Line waves are recently discovered wave entities that are localized along two directions, and therefore can be viewed as the one-dimensional counterpart of surface waves. These waves can be supported at discontinuities of the surface reactance and/or resistance of low-dimensional materials such as metasurfaces or graphene. Here, a broader class of non-Hermitian surface-impedance junctions is studied that can support coupled line waves, and allows investigating different one-dimensional waveguiding mechanisms in a unified framework. It is theoretically demonstrated that, under parity-time-symmetry conditions, exceptional points can occur in a truly flat-optics scenario, hence endowing these waveguiding systems with the attractive features of both line-wave and exceptional-point physics, and shedding further light on the phase transitions existing in these systems. 
It is also shown that the required surface-impedance parameters are compatible with those attainable with typical models of photoexcited graphene metasurfaces at terahertz frequencies. Besides providing additional understanding in the physics of line waves, which is still in its infancy, these results pave the way to  intriguing developments in the largely unexplored field of non-Hermitian flat optics, with possible applications ranging from sensing to lasing and on-chip optical signal processing.
\end{abstract}

\maketitle

\section{Introduction}
The dissipation and amplification of waves are fundamental physical processes of vast importance in science and technology. While the study of these phenomena has a long history, the seminal work by Bender and co-workers \cite{Bender:1998rs} on {\em parity time}  (${\cal PT}$) symmetry in quantum mechanics has sparked, over the past two decades, a sustained interest in the 
wave-physics of non-Hermitian systems \cite{El-Ganainy:2018nh}, opening intriguing avenues in their fundamental understanding and control.
Within this framework, ``exceptional points'' (EPs), i.e., spectral degeneracies characterized by the coalescence of two (or more) eigenvalues and associated eigenfunctions \cite{Heiss:2012tp}, are among the most ubiquitous and fascinating manifestations. 
In optics and photonics, EPs have been observed in a plethora of different waveguiding scenarios, including coupled dielectric  
waveguides \cite{Guo:2009oo,Ruter:2010oo},
optical cavities \cite{Peng:2014pt,Chang:2014pt},
time-domain photonic lattices \cite{Regensburger:2012pt},
chains of scatterers \cite{Othman:2017ep},
photonic  \cite{Zhen:2015sr} and plasmonic \cite{Park:2020sb} crystals,
plasmonic and topological systems \cite{Gangaraj:2018tw,Shastri:2020di}  (see also Refs. \cite{Miri:2019ep,Parto:2020nh} for recent reviews).
In essence, the phase transition occurring at an EP, and the associated square-root branch-point structure of the eigenvalue surfaces in parameter space, enables a variety of exotic responses and effects, and represents a ``sweet spot'' of operation in several application scenarios, including
lasing \cite{Feng:2014sm,Hodaei:2014pt}, 
unidirectional invisibility \cite{Ge:2012cr,Sounas:2015uc,Monticone:2016pt},
sensing \cite{Wiersig:2020ro}, slow light \cite{Goldzak:2018ls}, and
asymmetric mode conversion \cite{Doppler:2016de,Hassan:2017de,Khurgin:2020ee}.

Another waveguiding scenario of significant recent interest is the case of ``line waves'' (LWs), which exhibit both in-plane and out-of-plane confinement and can be supported by metasurfaces at planar discontinuities of the surface reactance \cite{Horsley:2014od,Bisharat:2017ge} or resistance \cite{Moccia:2020lw}. This enables a truly one-dimensional (1-D) energy flow, 
with a wealth of attractive features in terms of singular field enhancement, near-field structure, chiral coupling and topological-like robustness, 
which can be tailored and reconfigured by resorting to graphene \cite{Bisharat:2018ml} and/or gain-based  \cite{Moccia:2020lw} implementations. These characteristics are
of great interest for ``flat'' optics and photonics scenarios, with potential applications ranging from subdiffractive sensing and near-field imaging to optical and quantum computing. 

Here, we show that EPs can also occur for coupled LWs supported by non-Hermitian metasurfaces, endowing these emerging waveguiding systems with all the intriguing and useful properties of such spectral degeneracies. In particular, we focus on certain aspects that are distinctive of LW physics, and differ from typical non-Hermitian coupled-waveguide systems that are known to exhibit EPs. Furthermore, we illustrate how the phase transition at an EP embraces the known  classes of conventional \cite{Bisharat:2017ge} and non-Hermitian \cite{Moccia:2020lw} LWs, and we offer additional physical insight into how different 1-D waveguiding mechanisms can be continuously transformed into one another.

Accordingly, the rest of the paper is laid out as follows. In Sec. \ref{Sec:Problem}, we outline the problem formulation. In Sec. \ref{Sec:Results}, we present and discuss a body of representative results, including source-free (modal) and source-excited configurations, and relevant parametric studies. In Sec. \ref{Sec:Implementation}, we explore a possible physical implementation in terms of photoexcited graphene metasurfaces. Finally, in Sec. \ref{Sec:Conclusions} we draw some brief conclusions and discuss possible perspectives. 
 
 %
 \begin{figure}
 	\centering
 	\includegraphics[width=.6\linewidth]{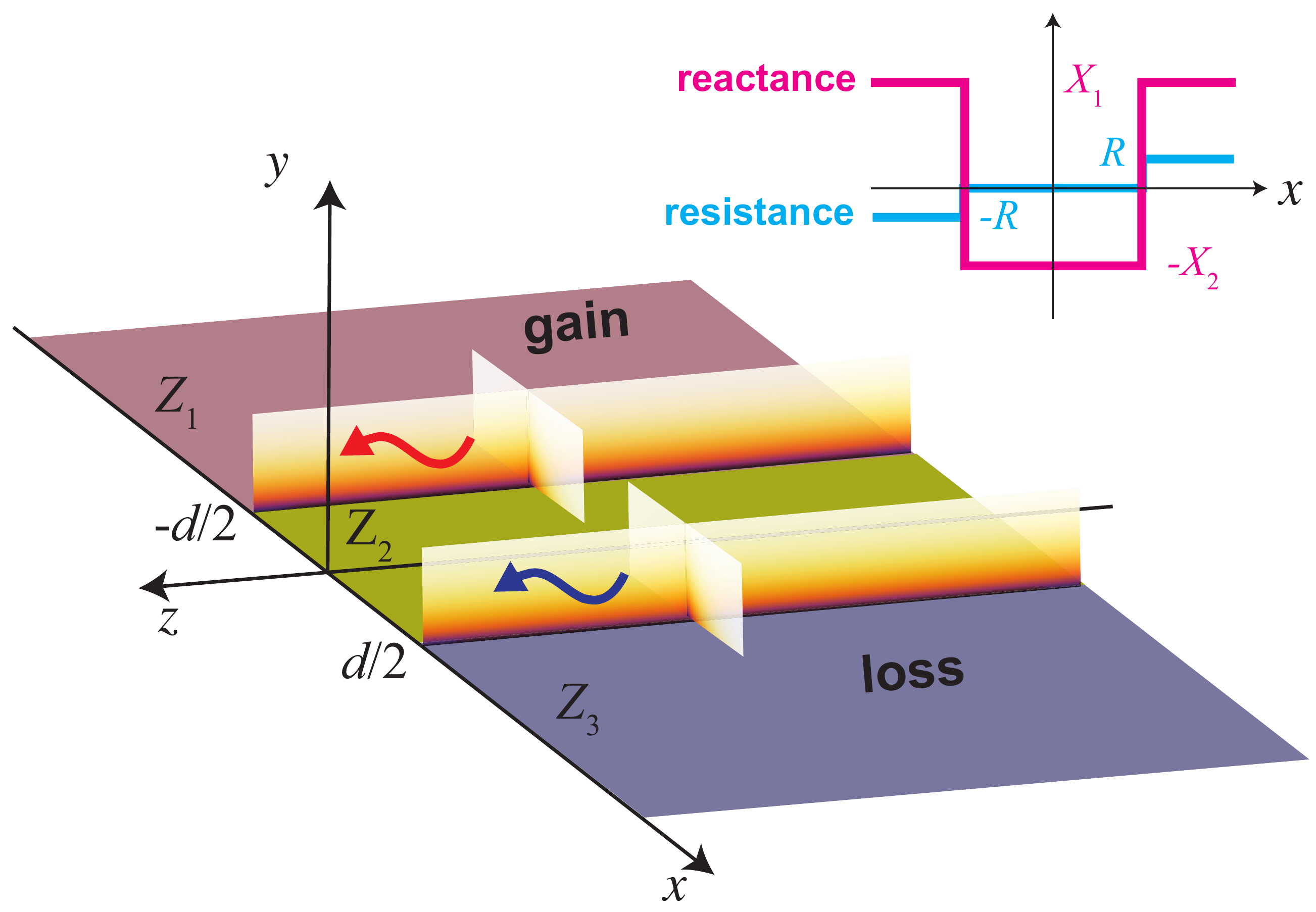}
 	\caption{Problem geometry: a non-Hermitian metasurface with a discontinuous profile of its surface resistance and reactance  (details in the text). The false-color maps qualitatively illustrate the unattenuated propagation along the $z$-direction and the in-plane and out-of-plane decay of the LWs supported at the two interfaces/discontinuities ($x=\pm d/2$).}
 	\label{Figure1}
 \end{figure}
 
 \section{Problem Geometry and Statement}
\label{Sec:Problem} 
Referring to Fig. \ref{Figure1} for schematic illustration, we consider a non-Hermitian metasurface, of infinite extent in the $x-z$ plane, characterized by an isotropic impedance boundary condition
${{{\bf E}_t} = Z\left( x \right){{\hat u}_y} \times {\bf H}}$ at $y = 0$, with the subscript ``$_t$'' denoting the tangential component, and ${\hat u}_y$  a $y$-directed unit vector; this model is appropriate for many classes of artificial and natural low-dimensional materials.
The surface impedance features a double discontinuity which, for a suppressed $\exp(-i\omega t)$ time-harmonic convention, can be parameterized as
\beq
Z\left( x \right) = \left\{ \begin{array}{l}
	{Z_1} =  - R + i{X_1},\quad\quad\quad x <  - d/2,\\
	{Z_2} =  -i{X_2},\quad\quad \quad \quad  - d/2 < x < d/2,\\
	{Z_3} =-Z_1^*= R + i{X_1}, \quad x > d/2.
	\label{eq:Zs}
\end{array} \right.
\eeq
In what follows, we start by assuming positive values for $X_1, X_2, R$,  which implies that the three regions are capacitive gainy, purely inductive, and capacitive lossy, respectively.
However, we will show that similar results can be obtained for negative values as well. 
It is readily verified that the surface-impedance distribution in (\ref{eq:Zs}) satisfies the ${\cal PT}$-symmetry condition
\beq
Z\left( { - x} \right) =  - {Z^*}\left( x \right).
\label{eq:PTS}
\eeq
Intuitively, as an effect of the capacitive/inductive discontinuities, the two interfaces $x=\pm d/2$ may support LWs similar to those studied in Refs. \cite{Horsley:2014od,Bisharat:2017ge}, exponentially bound along the $x$ and $y$ directions, and propagating along the $z$ direction with attenuation ($x=d/2$) or amplification ($x=-d/2$) due to the presence of loss and gain, respectively. In other words, when the interfaces are well separated, the LWs are independently supported by oppositely signed reactances at each interface, and are analogous to surface modes at $\varepsilon$-negative/$\mu$-negative interfaces \cite{Alu:2003pa}, arising from the dual nature of the two media. The coupling between these LWs can then be controlled by varying the
interface separation $d$, and therefore this configuration resembles typical non-Hermitian scenarios featuring coupled waveguides with gain/loss spatial modulation, which are known to exhibit EPs \cite{Miri:2019ep}. However, we stress that this extension is far from straightforward, as there are a number of important aspects that are distinctive of the LW physics and deserve a deeper investigation. First, in view of the complicate dispersion relationship of LWs \cite{Kong:2019at}, their {\em a priori} existence and bound character cannot be taken for granted, as they depend on the delicate interplay between the surface reactance and resistance. For instance, the addition of gain/loss to a purely reactive discontinuity may induce a {\em leaky} (radiative) character \cite{Moccia:2020lw}. Second, the ideal (discontinuous) model of the surface impedance, which works reasonably well for single LWs \cite{Horsley:2014od,Bisharat:2017ge,Moccia:2020lw}, is expected to break down in the strong coupling regime, and some regularization is needed in order to avoid unphysical behaviors. Last, but not least, it is crucial to explore the practical feasibility of such a platform, in terms of physical models of metasurfaces featuring realistic levels of gain. Our study here, motivated by the above issues, bridges the two insofar known LW scenarios of purely reactive (capacitive/inductive) \cite{Horsley:2014od,Bisharat:2017ge} and ${\cal PT}$-symmetric \cite{Moccia:2020lw} junctions. In this latter case (recovered in the limit $d\rightarrow 0$), LWs are supported by the opposite resistances (gain/loss) at the interface. The considered structure, therefore, allows studying the two known classes of LWs in a unified framework, and allows clarifying the transition between these two regimes mediated by the occurrence of an EP.

%
\begin{figure}
	\centering
	\includegraphics[width=.6\linewidth]{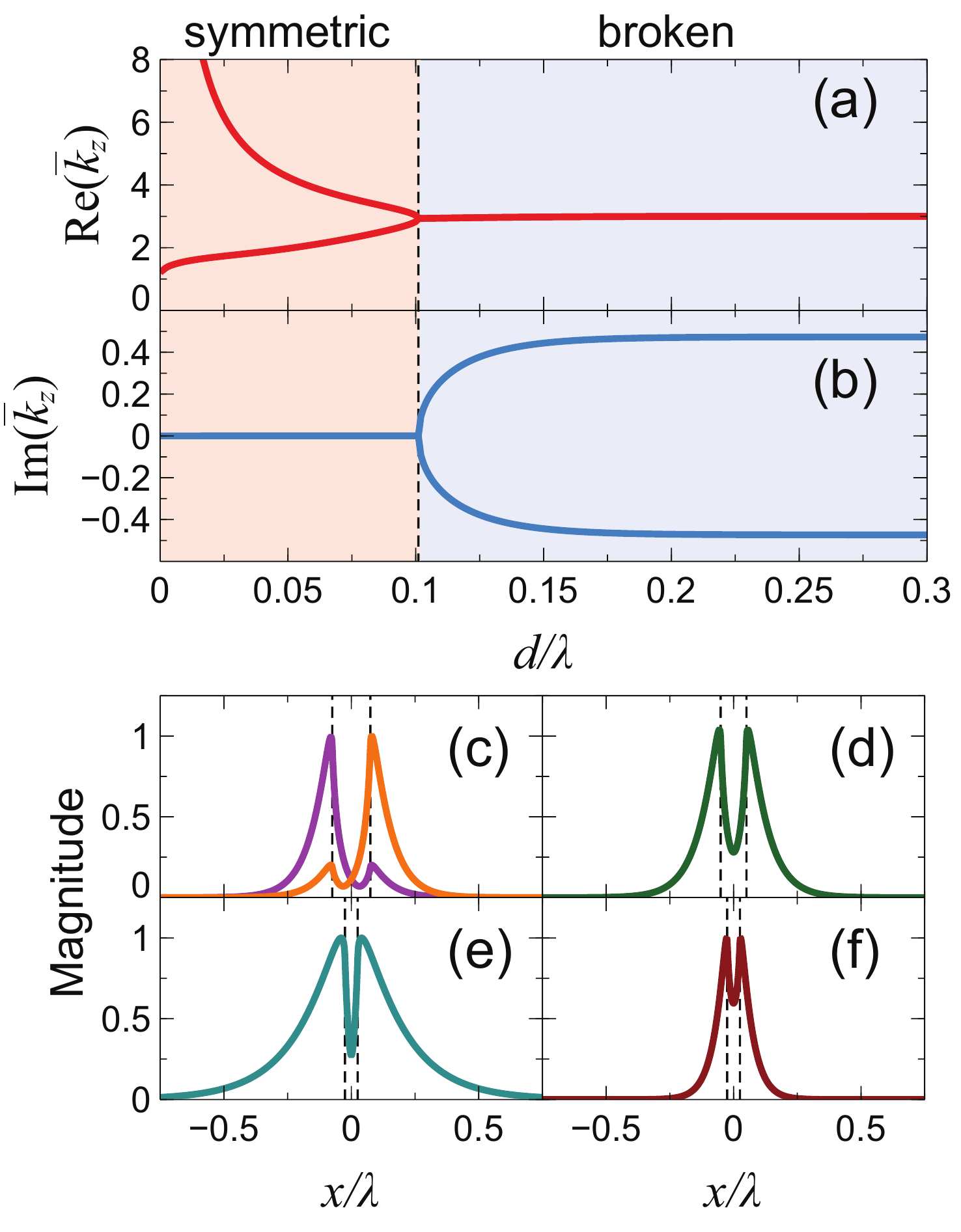}
	\caption{Eigenmode analysis. (a), (b) Real and imaginary part, respectively, of mode index, as a function of $d/\lambda$,  for ${\bar X_1}=0.7$, ${\bar X_2}=0.2$, ${\bar R}=0.1$; the symmetric and broken phases are highlighted with different color shades. (c) Modal-field profiles ($|E_z|$) at $d=0.15\lambda$ (purple: ${\bar k}_z=2.980-i0.444$; orange: ${\bar k}_z=2.980+i0.444$). (d) Modal-field profile at $d=0.101\lambda$ (EP, ${\bar k}_z=2.931$). (e), (f) Modal-field profiles at $d=0.05\lambda$ (${\bar k}_z=1.976$ and ${\bar k}_z=4.255$, respectively). Modes are normalized with respect to their maximum magnitude; the vertical dashed lines indicate the position of the interfaces at $x=\pm d/2$.}
	\label{Figure2}
\end{figure}

%
\begin{figure*}
	\centering
	\includegraphics[width=\linewidth]{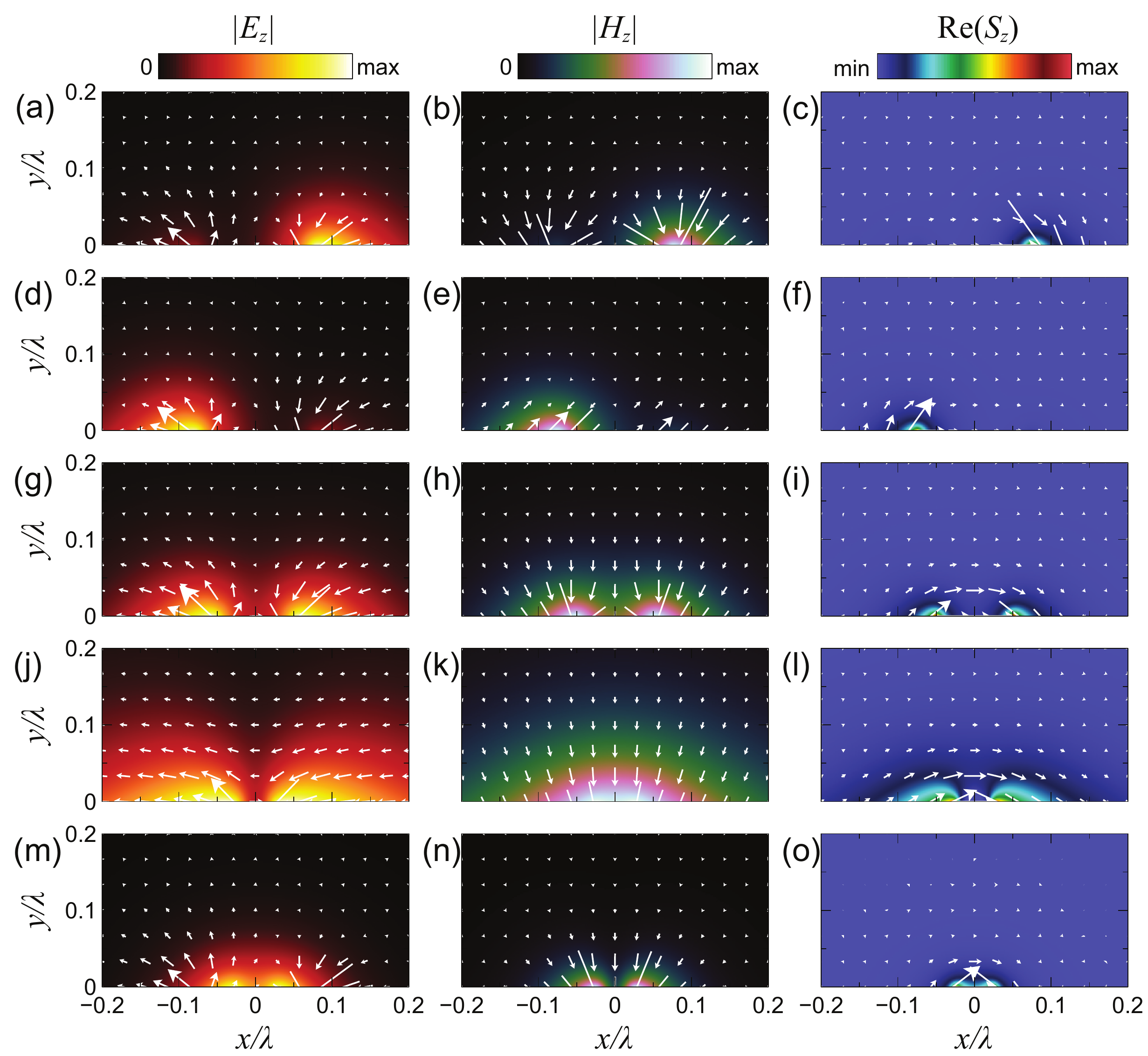}
	\caption{(a), (b), (c) Numerically computed electric and magnetic modals fields, and power flow (real part of Poynting vector ${\bf S}$), respectively, for 
		${\bar X_1}=0.7$, ${\bar X_2}=0.2$, ${\bar R}=0.1$ and $d=0.15\lambda$ (${\bar k}_z=2.980+i0.444$).
		(d), (e), (f) Same as above, but ${\bar k}_z=2.980-i0.444$.
		(g), (h), (i) Same as above, but at $d=0.101\lambda$ (EP, ${\bar k}_z=2.931$).
		(j), (k), (l) Same as above, but at $d=0.05\lambda$ (${\bar k}_z=1.976$).
		(m), (n), (o) Same as above, but ${\bar k}_z=4.255$.
		The $z$-components are shown as false-color maps, whereas the corresponding $x$ and $y$ components are shown as vector plots.}
	\label{Figure3}
\end{figure*}

\section{Representative Results}
\label{Sec:Results}

\subsection{Eigenmode Analysis}

Referring to Appendix \ref{Sec:AppA} for details on the modeling aspects, we start with the eigenmode (i.e., source-free) analysis of a representative parameter configuration featuring ${\bar X}_1=0.7$, ${\bar X}_2=0.2$ and ${\bar R}=0.1$, with the overbar denoting a normalization with respect to the vacuum characteristic impedance $\eta=\sqrt{\mu_0/\varepsilon_0}\approx377\Omega$. Figures \ref{Figure2}a and \ref{Figure2}b show the numerically computed mode index (real and imaginary part, respectively) ${\bar k}_z=k_z/k$, as a function of the electrical thickness $d/\lambda$ of the central region; here, $k_z$ and $k=\omega/c=2\pi/\lambda$ are the propagation constant along the $z$-direction and the vacuum wavenumber, respectively (with $c$ and $\lambda$ denoting the vacuum wavespeed and wavelength, respectively). A phenomenon of ``spontaneous symmetry breaking'' is clearly observed, with a transition between a ``symmetric'' and a ``broken'' phase through an EP. This behavior, characterized by a square-root branch-point bifurcation of the eigensolutions around the EP, is typical of ${\cal PT}$-symmetric coupled optical systems \cite{Ruter:2010oo}, but has never been predicted or observed before in the context of LWs. As expected, for sufficiently large values of the interface separation $d$, two complex-conjugate solutions are observed, with the corresponding modes (Fig. \ref{Figure2}c) mostly localized at the interfaces $x=\pm d/2$; essentially, these modes are the conventional (capacitive/inductive) LW solutions in Ref. \cite{Horsley:2014od,Bisharat:2017ge} with attenuation ($x=d/2$) or amplification ($x=-d/2$), only weakly interacting. By reducing the separation, the modal coupling increases, and the amplification/attenuation effects progressively compensate. At a critical value of the separation ($d_{EP}=0.101\lambda$ in this example), the two solutions coalesce, yielding an EP, with a real-valued propagation constant and a symmetric modal distribution (Fig. \ref{Figure2}d). For smaller values of the separation, two real-valued branches are observed, with symmetric modal distributions (Figs. \ref{Figure2}e and \ref{Figure2}f). 

%
\begin{figure}
	\centering
	\includegraphics[width=.6\linewidth]{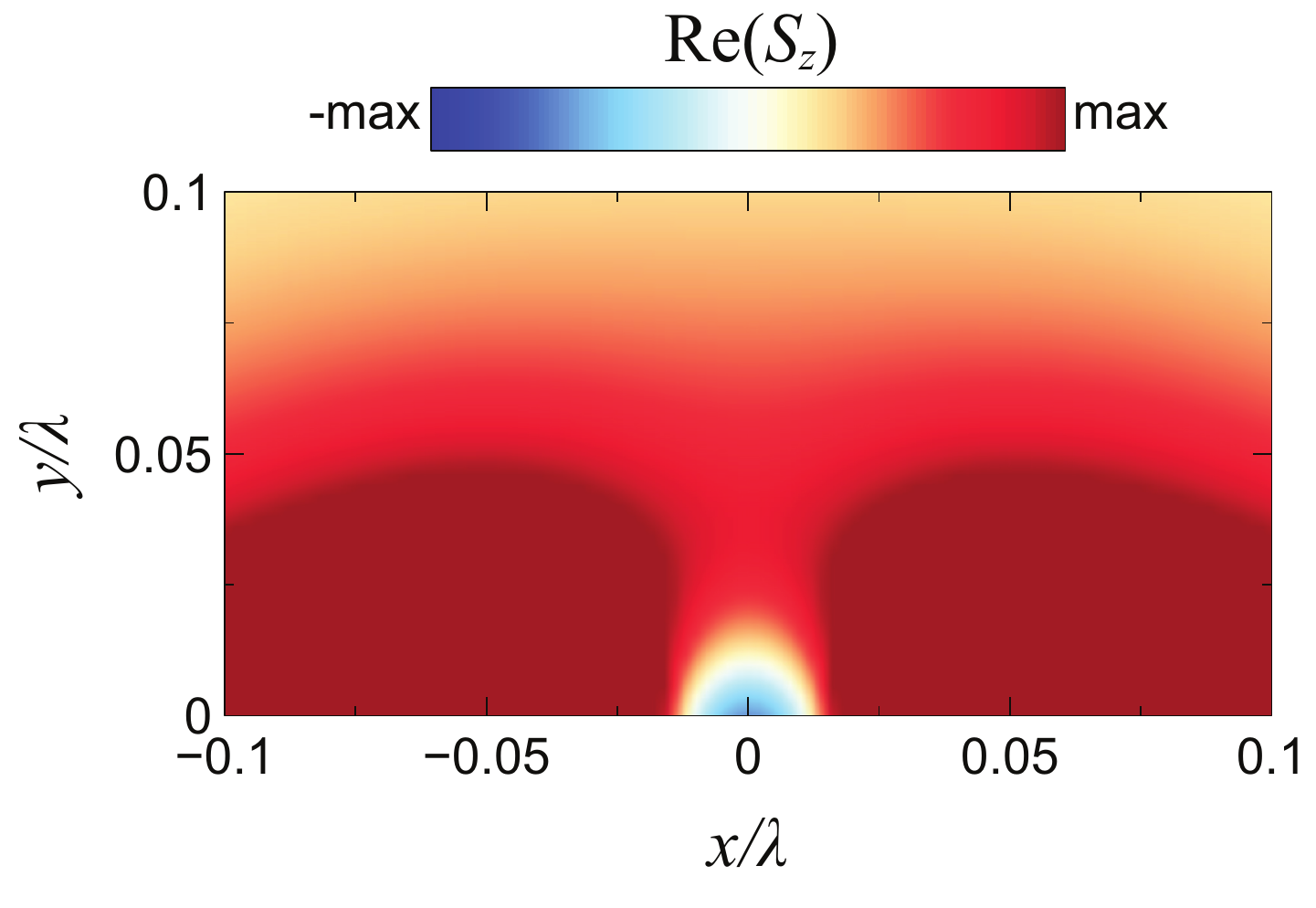}
	\caption{As in Fig. \ref{Figure3}l ($d=0.05\lambda$, ${\bar k}_z=1.976$), but with magnified view and diverging colormap, so as to highlight the negative values around $x=y=0$.}
	\label{Figure4}
\end{figure}

The complete modal-field distributions of these representative cases are shown in Fig. \ref{Figure3}.  Specifically, we show (left to right) the electric and magnetic fields (in a plane orthogonal to the propagation direction) as well as the active power density (real part of the Poynting vector ${\bf S}$), for (top to bottom) representative parameter configurations in the broken phase ($d=0.15\lambda$, ${\bar k}_z=2.980+i0.444$, ${\bar k}_z=2.980-i0.444$), at the EP ($d=0.101\lambda$, ${\bar k}_z=2.931$), and in the symmetric phase ($d=0.05\lambda$, (${\bar k}_z=1.976$, ${\bar k}_z=4.255$).
First, as for conventional \cite{Bisharat:2017ge} and non-Hermitian \cite{Moccia:2020lw} LWs, we observe the generally {\em hybrid} character of the fields, with nonzero electric and magnetic components along the propagation direction. Second, we note that the transverse fields exhibit some handedness properties that can be associated with spin-momentum locking (see also our discussion in Sec. \ref{Sec:Source} below). Finally, we highlight the transverse power flow from the gain to the loss region. In the broken phase (Figs. \ref{Figure3}c and \ref{Figure3}f), this flow is not sufficient to balance the gain and loss effects, resulting in a net attenuation or amplification along the propagation direction. Conversely, in the symmetric phase (Figs. \ref{Figure3}l and \ref{Figure3}o), a perfect compensation is attained, and the fields propagate without attenuation or amplification. Our results show that, also in the context of LWs, the EP is exactly where, as the coupling increases, such transverse power flow becomes sufficiently strong to perfectly balance dissipation and gain, marking the transition to a different regime of LW propagation.
This transverse power flow is inherent of non-Hermitian configurations, and has also been observed in previous studies on coupled optical waveguides featuring gain and loss \cite{Miri:2019ao}. Interestingly, it has been shown that  it can give rise to anomalous optical forces, controllable via the gain/loss level, of potential interest for microfluidics and micro-optomechanical applications \cite{Miri:2019ao}. 

Interestingly, there are some regions where $\mbox{Re}\left(S_z\right)$ is locally negative. For instance, although this is somehow masked by the colorscale, for one mode in the symmetric phase ($d=0.05\lambda$, ${\bar k}_z=1.976$; Fig. \ref{Figure3}l) the longitudinal power flow locally exhibits negative values around $x=y=0$. This is better highlighted in Fig. \ref{Figure4}, with a magnified view and a diverging (and suitably saturated) colorscale. Nevertheless, in all examples considered, the net longitudinal power flow [integral of $\mbox{Re}\left(S_z\right)$ across the section] is always positive, i.e., directed along the propagation direction.

%
\begin{figure}
	\centering
	\includegraphics[width=.7\linewidth]{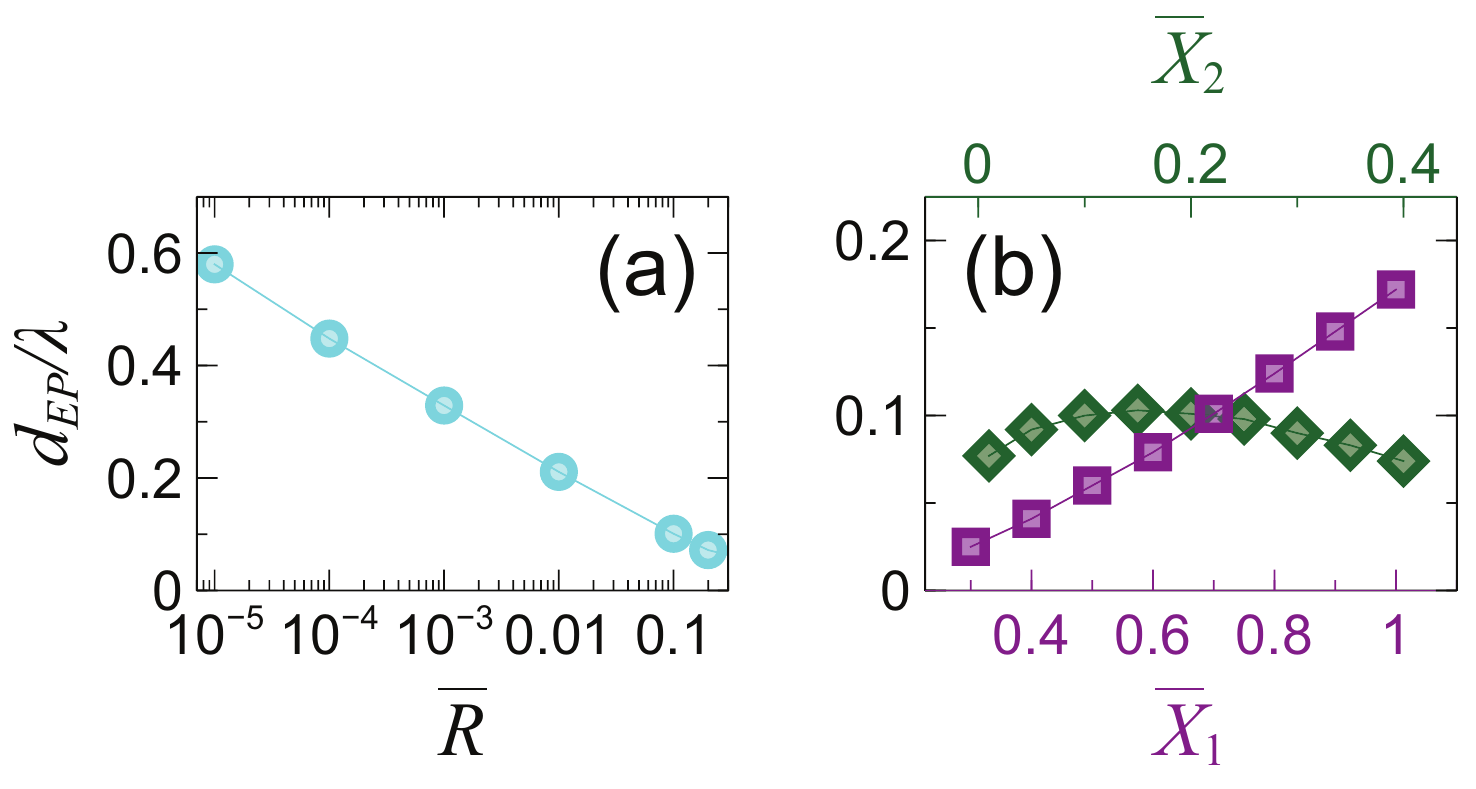}
	\caption{(a) Critical electrical separation between the metasurface interfaces to obtain an EP ($d_{EP}/\lambda$) as a function of normalized gain/loss parameter ${\bar R}$, for ${\bar X_1}=0.7$, ${\bar X_2}=0.2$. 
		(b) Same as above (with ${\bar R}=0.1$), but as a function of ${\bar X_1}$ (purple squares, bottom axis) for ${\bar X_2}=0.2$,  and ${\bar X_2}$ (green diamonds, top axis) for ${\bar X_1}=0.7$.}
	\label{Figure5}
\end{figure}

\subsection{Parametric Studies}
Figure \ref{Figure5} illustrates the dependence of the EP on the surface-impedance parameters. Specifically, from Fig. \ref{Figure5}a, we observe that the critical interface separation $d_{EP}$ varies logarithmically with the gain/loss parameter ${\bar R}$. As can be expected, the EP asymptotically disappears in the Hermitian limit ${\bar R}\rightarrow 0$. Interestingly, as can be observed from  Fig. \ref{Figure5}b, the EP also depends on the reactance parameters ${\bar X}_1$ and ${\bar X}_2$. In particular, $d_{EP}$ progressively decreases for decreasing values of ${\bar X}_1$, as an effect of the increasing localization of the LW supported at the two interfaces.

%
\begin{figure}
	\centering
	\includegraphics[width=.6\linewidth]{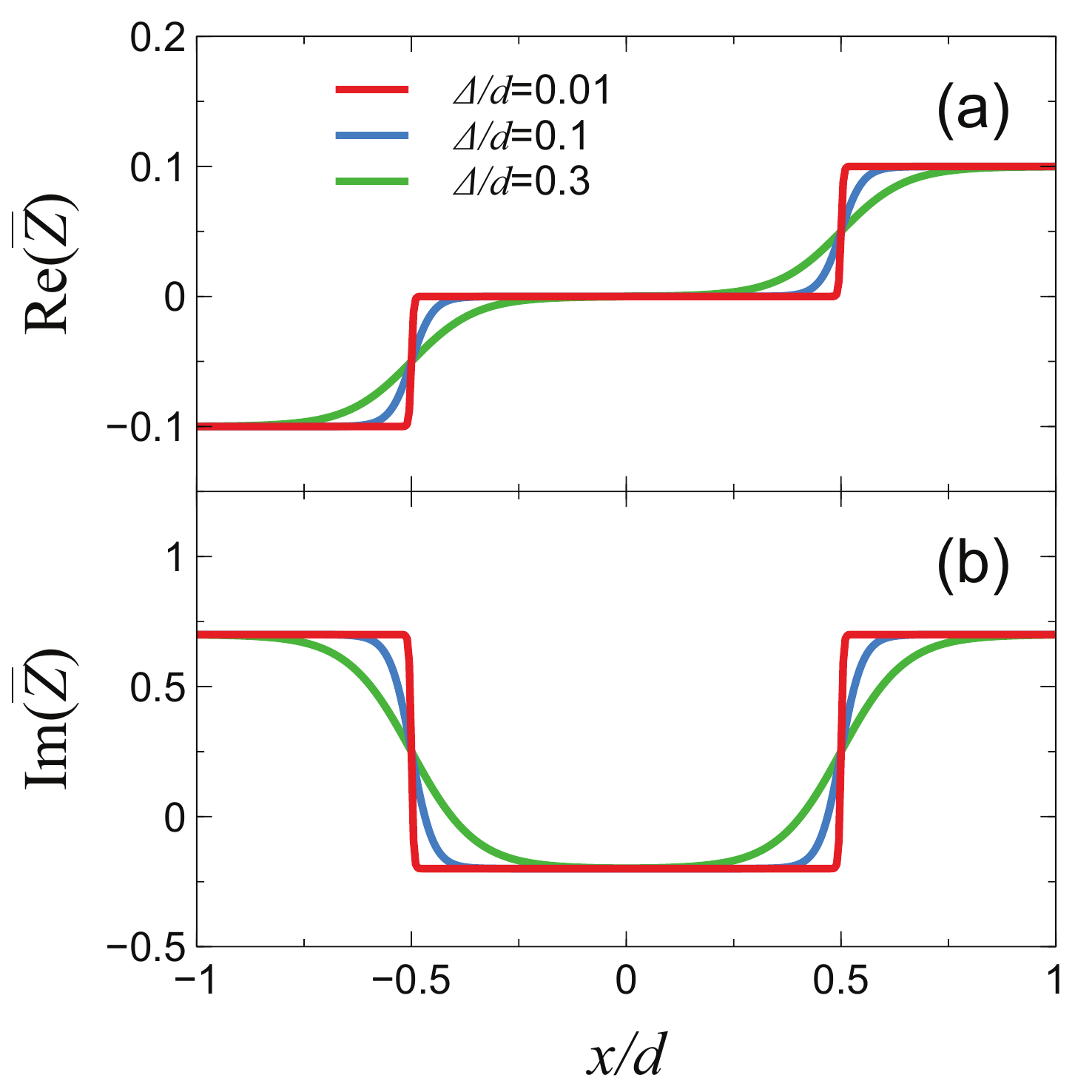}
	\caption{(a), (b) Real and imaginay part, respectively, of spatial profile of the continuous impedance model in (\ref{eq:Ztanh}) (normalized by the vacuum characteristic impedance), for 
		${\bar X_1}=0.7$, ${\bar X_2}=0.2$, ${\bar R}=0.1$ and different values of the steepness parameter $\Delta/d$.}
	\label{Figure6}
\end{figure}

%
\begin{figure}
	\centering
	\includegraphics[width=.6\linewidth]{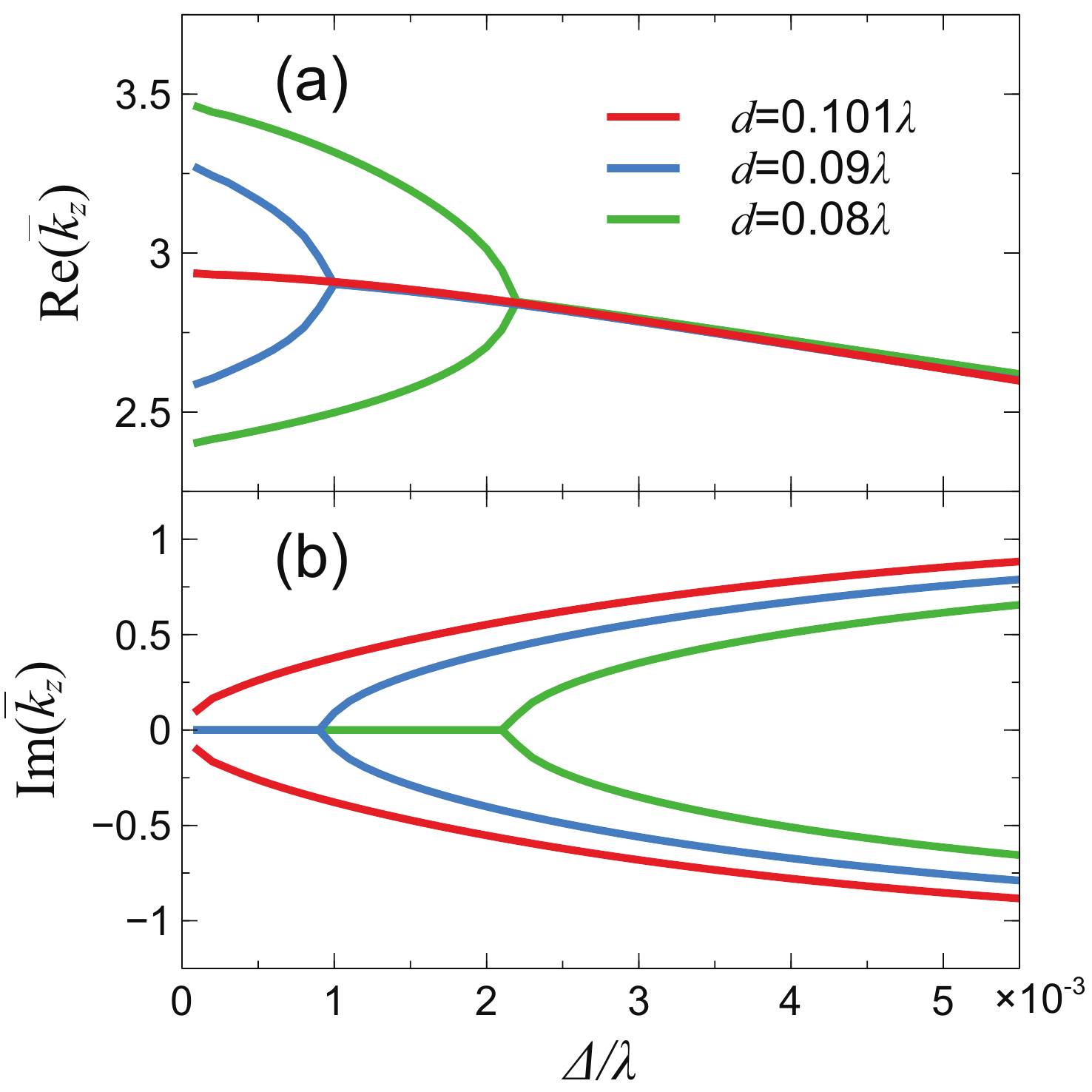}
	\caption{(a), (b) Real and imaginary part, respectively, of mode index, for the continuous surface-impedance profile in (\ref{eq:Ztanh}) (with ${\bar X_1}=0.7$, ${\bar X_2}=0.2$, ${\bar R}=0.1$)
		as a function of the steepness parameter $\Delta/\lambda$,  for different values of the electrical separation $d/\lambda$.}
	\label{Figure7}
\end{figure}

From Fig. \ref{Figure2}a, we also observe that, in the limit $d\rightarrow 0$, the mode index grows indefinitely for the upper branch whereas, for the lower one, it approaches the value found in Ref. \cite{Moccia:2020lw} (${\bar k}_z=1.213$ in this example) for non-Hermitian LWs in capacitive, ${\cal PT}$-symmetric single discontinuities. 
Intuitively, the mode-index divergence can be explained by noting that the upper branch pertains to a mode supported by the reactance discontinuities, whose localization progressively increases as $d$ decreases (a higher degree of localization corresponds to a larger imaginary transverse wavenumber and, therefore, a larger longitudinal wavenumber ${\bar k}_z$). This unphysical behavior is attributable to the breakdown of the ideal (discontinuous) model of the surface impedance.
To better understand this aspect, we assume a more realistic model featuring a smooth transition
\beq
Z\left(x\right)\!=\!\left\{\begin{array}{l}
	Z_{1}\!-\!\displaystyle{\frac{\left(Z_{1}\!-\!Z_{2}\right)}{2}}\left[1\!+\!\tanh \left(\frac{x\!+\!d/2}{\Delta}\right)\right], \!\quad x<0, \\
	~\\
	Z_{2}\!-\!\displaystyle{\frac{\left(Z_{2}\!-\!Z_{3}\right)}{2}}\left[1\!+\!\tanh \left(\frac{x\!-\!d/2}{\Delta}\right)\right], \!\quad x>0,
\end{array}\right.
\label{eq:Ztanh}
\eeq
with the parameter $\Delta$ controlling the steepness. For the same impedance parameters as in Fig. \ref{Figure2}, Fig. \ref{Figure6} shows some representative smoothed profiles for different values of the steepness parameter $\Delta/d$. Figure \ref{Figure7} shows the effects of the smoothing in the mode index, for selected values of the interface separation $d$ around the EP.  As can be observed, for a given value of $d$, the EP depends on the steepness parameter $\Delta$. For instance, for $d=0.101\lambda$, the EP is attained in the asymptotic limit $\Delta\rightarrow 0$ (i.e., discontinuous profile), consistently with the results in Fig. \ref{Figure2}. As $d$ decreases, the EP is attained at {\em finite} (and progressively larger) values of $\Delta$. It is also apparent that, in the symmetric phase, the growth of the upper branch of the mode index is inherently limited when finite values of $\Delta$ are assumed.
To sum up, the assumption of a continuous surface-impedance profile removes the aforementioned unphysical behavior.
This is analogous to the unphysical divergent behavior of surface waves in different scenarios, which can similarly be regularized by considering smoother transitions or spatial dispersion \cite{Monticone:2020at,Gangaraj:2018tw}. 
We stress, however, that even after proper regularization, this LW solution retains a moderately large propagation constant corresponding to a highly localized mode, which may be useful for subwavelength sensing and near-field imaging applications.

%
\begin{figure}
	\centering
	\includegraphics[width=.6\linewidth]{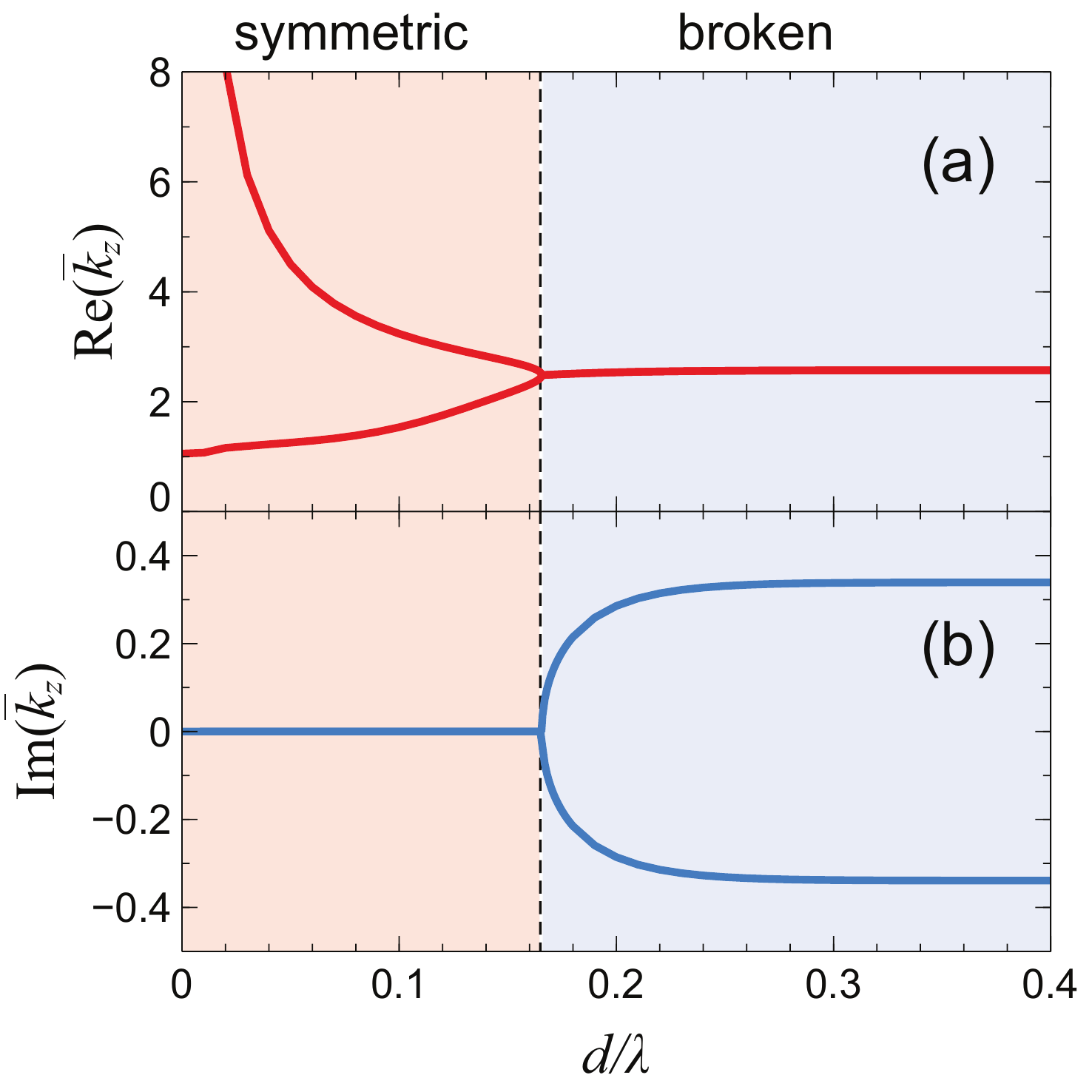}
	\caption{
		(a), (b) Real and imaginary part, respectively, of mode index, as a function of $d/\lambda$,  for ${\bar X_1}=-0.7$, ${\bar X_2}=-2$, ${\bar R}=0.1$; the symmetric and broken phases are highlighted with different color shades.	
	}
	\label{Figure8}
\end{figure}

%
\begin{figure*}
	\centering
	\includegraphics[width=\linewidth]{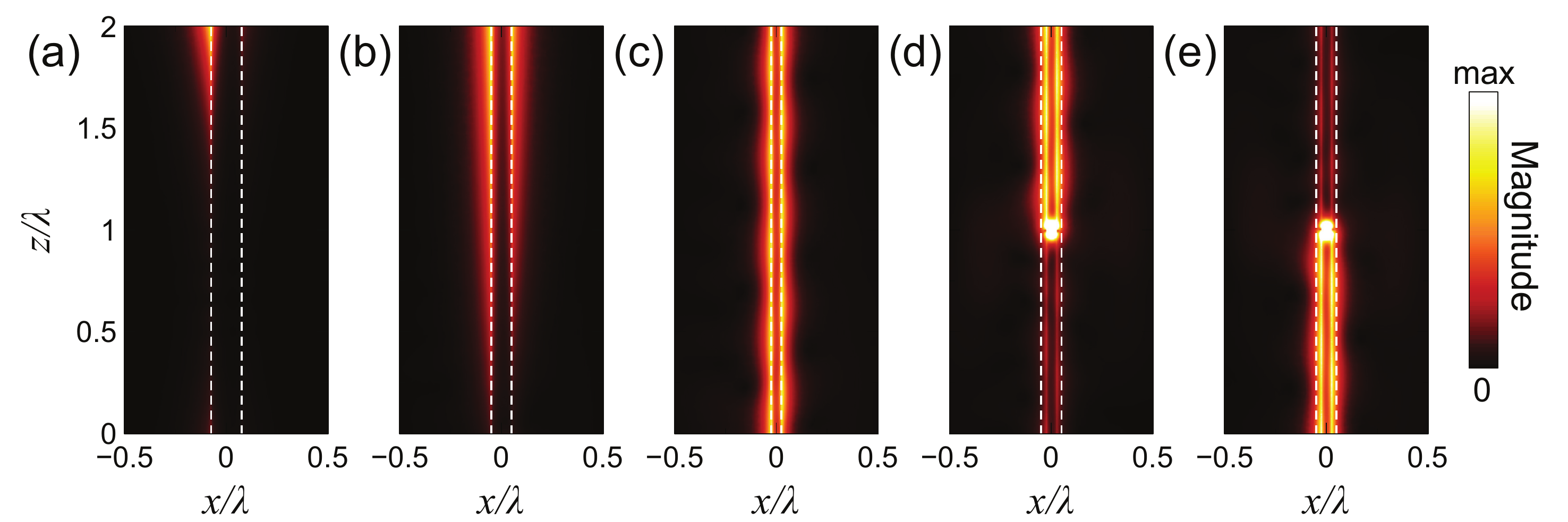}
	\caption{(a), (b), (c) Numerically computed in-plane ($y = 0.001\lambda$) electric-field magnitude maps in false color scale for ${\bar X_1}=0.7$, ${\bar X_2}=0.2$, ${\bar R}=0.1$, and $d=0.15\lambda$, $d=0.101\lambda$, $d=0.05\lambda$, respectively. The field is excited by a $y$-directed elementary electric dipole placed at $x=0$, $y=0.02\lambda$, and $z = -0.5\lambda$ (outside the domains visualized). (d), (e) Same as above but for $d=0.05\lambda$ and excitation by 
		circularly polarized elementary sources with right and left handedness,  respectively, placed at $x=0$, $y=0.02\lambda$, $z=\lambda$.}
	\label{Figure9}
\end{figure*}

%
\begin{figure}
	\centering
	\includegraphics[width=.6\linewidth]{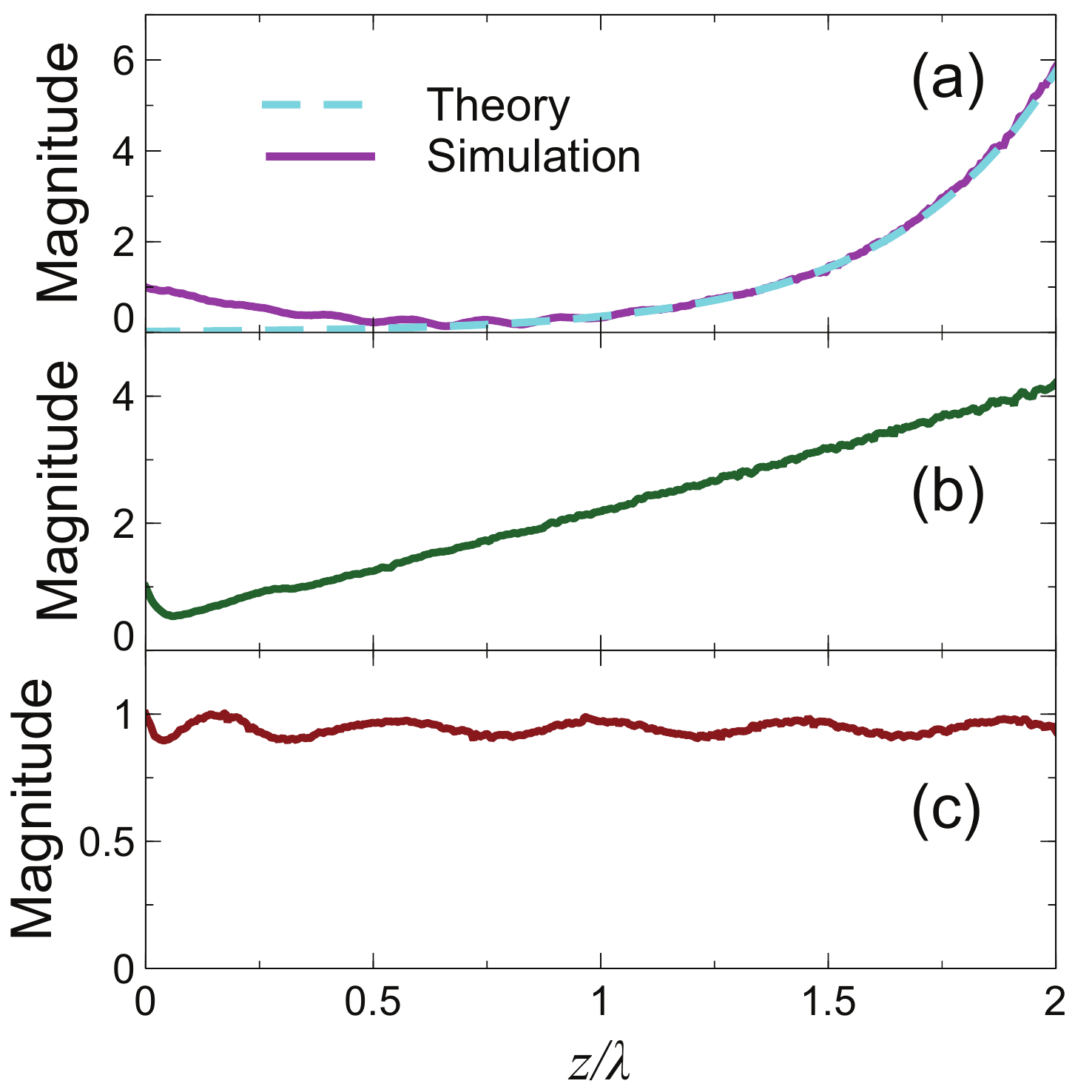}
	\caption{(a), (b), (c) Longitudinal cuts (at $x=0$) of the field maps in Figs. \ref{Figure9}a, \ref{Figure9}b, \ref{Figure9}c, respectively. The theoretical curve in panel (a) is based on the mode index ${\bar k}_z=2.980-i0.444$ of the amplified solution.}
	\label{Figure10}
\end{figure}

As previously mentioned, EPs can also occur in dual surface-impedance configurations ($X_1,X_2<0$, i.e., inductive gainy, purely capacitive, and inductive lossy).
Figure \ref{Figure8} illustrates some representative results for one such parameter configuration featuring  ${\bar Z}_1=-0.1-i0.7$, 
${\bar Z}_2=2i$, ${\bar Z}_3=-{\bar Z}_1^*=0.1-i0.7$.  
Also in this case we observe a phase transition, with an EP at $d_{EP}/\lambda=0.165$. In the limit $d\rightarrow 0$, the modal divergence of the upper branch can be explained (and removed) using similar arguments as for the previous configuration.
However, by comparison with Fig. \ref{Figure2}a, we note an important difference, as the lower modal branch now tends toward the light line ${\bar k}_z=1$.
This is consistent with the theory in Ref. \cite{Moccia:2020lw}, which predicts a leaky LWs at a ${\cal PT}$-symmetric interface with
$\left|{\bar Z}_l\right|=\left|{\bar Z}_r\right|<1$, and is one of the distinctive aspects of non-Hermitian LW physics.

%
\begin{figure}
	\centering
	\includegraphics[width=.6\linewidth]{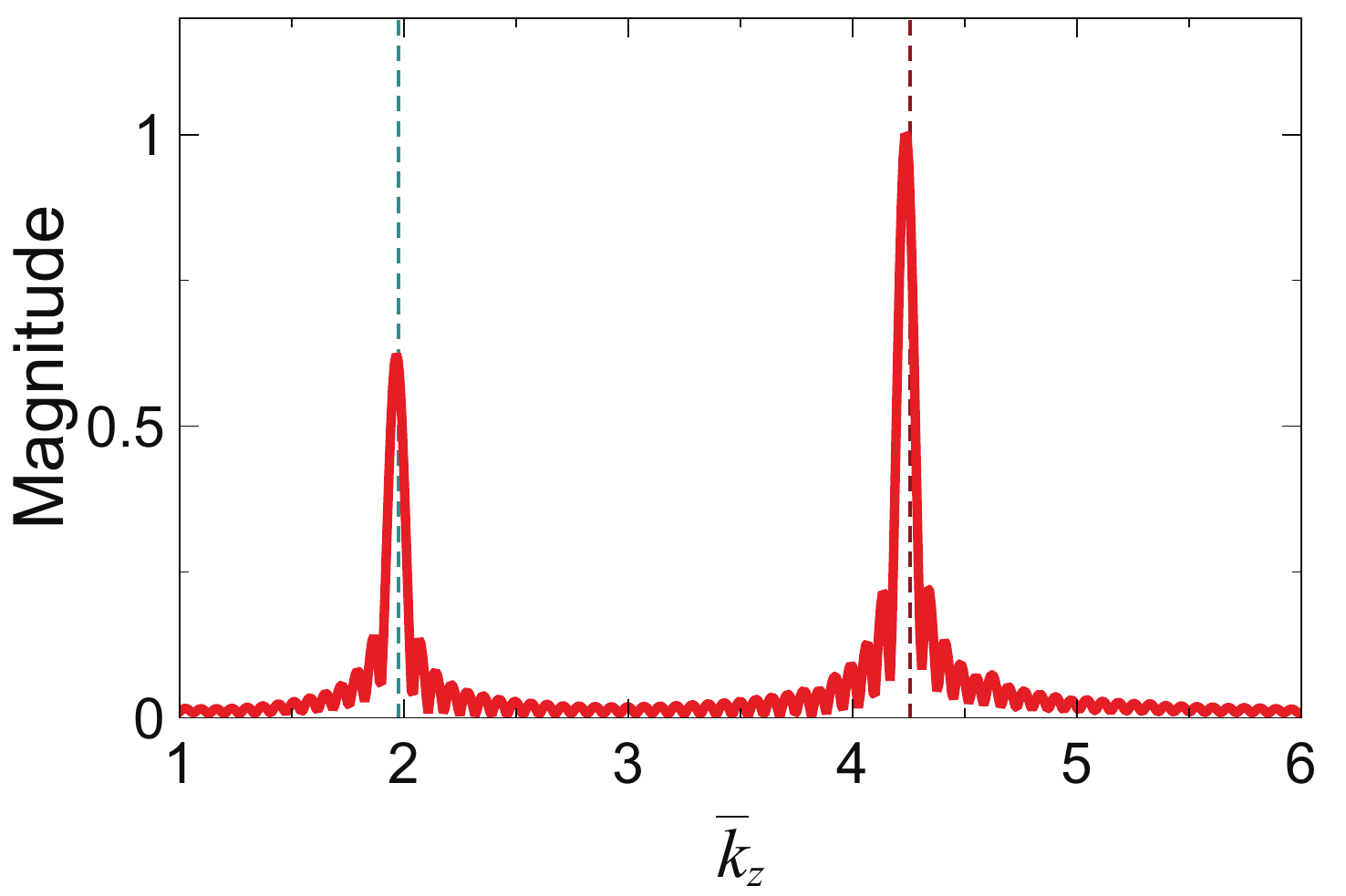}
	\caption{Spatial spectrum (Fourier transform magnitude) pertaining to the longitudinal cut in Figure \ref{Figure10}c. The vertical dashed lines indicate the nominal mode indices ${\bar k}_z=1.976$ and ${\bar k}_z=4.255$.}
	\label{Figure11}
\end{figure}

\subsection{Source-Excited Configurations}
\label{Sec:Source}
For the representative cases in Fig. \ref{Figure2},  the results of 3-D full-wave numerical simulations in the presence of elementary-dipole excitation are shown in Figs. \ref{Figure9}a-\ref{Figure9}c, with the corresponding cuts along the $z$-direction shown in Fig. \ref{Figure10}. 
Consistently with the eigenmode analysis, in the broken phase (Figs. \ref{Figure9}a), we observe a localization at the gainy interface $x=-d/2$ and an exponential growth along the $z$-direction (a LW is also present at the lossy interface, but is not visible in the figure as it is quickly attenuated). In Fig. \ref{Figure10}a, this response is also compared with the prediction from the eigenmode analysis (amplified modal branch with ${\bar k}_z=2.980-i0.444$). As can be observed, with the exception of an initial transition region (attributable to the space-wave effects of the dipole source), the agreement is excellent.

At the EP (Figs. \ref{Figure9}b and \ref{Figure10}b), we observe instead a transversely symmetric field magnitude distribution and the characteristic {\em linear} growth along the $z$-direction.
This is a well-known behavior, also known as a Jordan mode \cite{Longhi:2010ss,Graefe:2011pt,Shipman:2016ps,Hanson:2019ep}, which is not indicative of amplification (the mode index is real-valued), but rather stems from the interference between two coalescing eigenmodes with progressively larger amplitudes (see also the discussion in Ref. \cite{Gangaraj:2018tw}).

Finally, in the symmetric phase (Figs. \ref{Figure9}c and \ref{Figure10}c) we observe transverse symmetry and some longitudinal oscillations stemming from the interference of the two modes. This is more clearly visible in the spatial spectrum shown in Fig. \ref{Figure11}, from which we note two dominant peaks in close correspondence with the nominal mode indices (${\bar k}_z=1.976$ and ${\bar k}_z=4.255$) obtained from the eigenmode analysis. 
Interestingly, in the symmetric phase, the fields maintain the spin-momentum locking properties observed for conventional \cite{Bisharat:2017ge} and non-Hermitian \cite{Moccia:2020lw} LWs. Accordingly, as illustrated in Figure \ref{Figure9}d and \ref{Figure9}e, they exhibit chiral-coupling properties with respect to circularly polarized sources, i.e., unidirectional excitation and propagation with direction depending on the source handedness, which could be of significance in chiral quantum optics and valleytronics \cite{Lodahl:2017cq}.

%
\begin{figure}
	\centering
	\includegraphics[width=.6\linewidth]{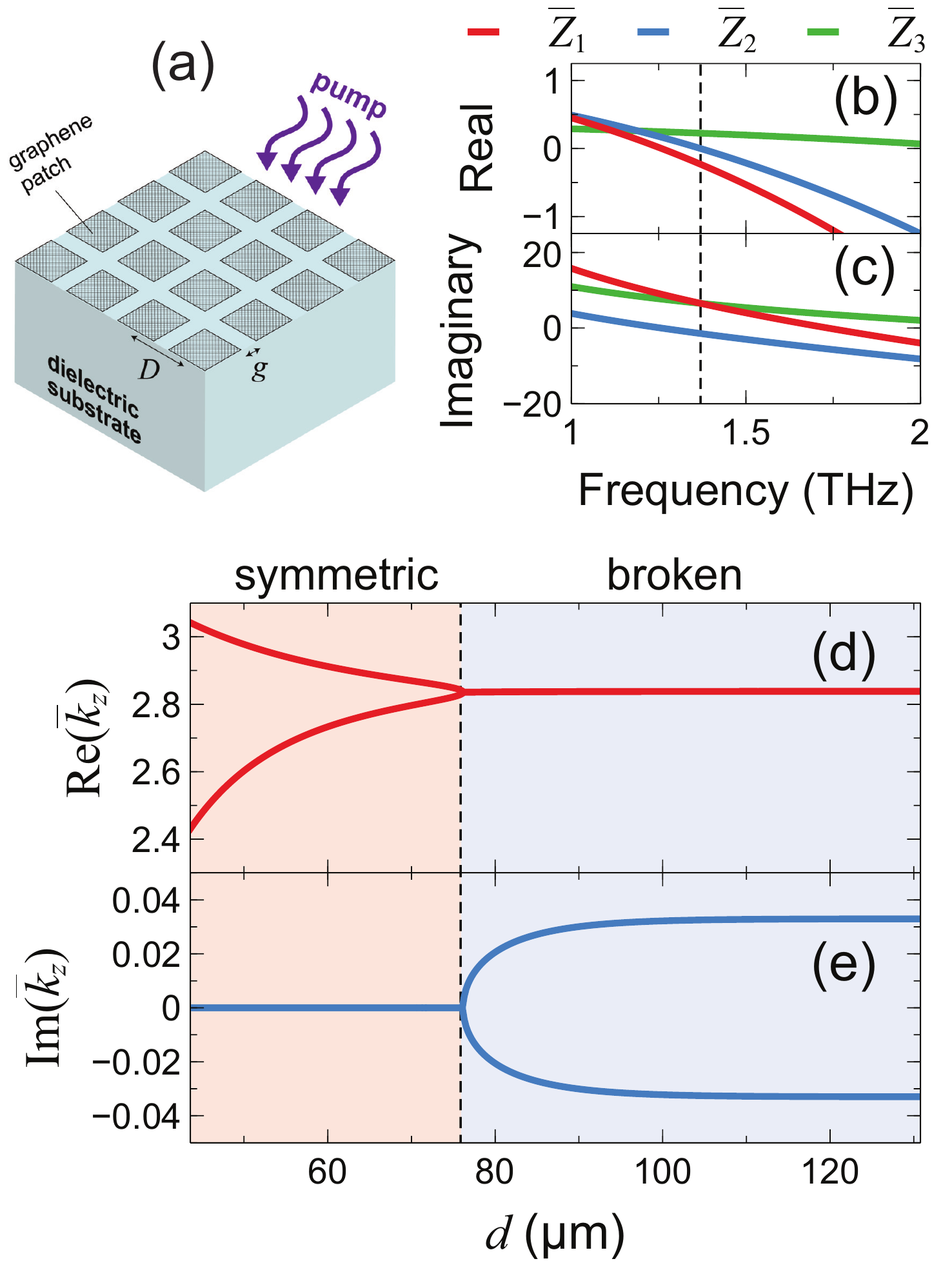}
	\caption{(a) Schematic of a photoexcited graphene-based metasurface.
		(b), (c) Real and imaginary part, respectively, of normalized surface impedances pertaining to the three sections of Fig. \ref{Figure1}, implemented as in panel (a), as a function of frequency (see Appendix \ref{Sec:AppB} for parameters and details). The dashed vertical line indicates the operational frequency (1.37 THz) at which the ${\cal PT}$-symmetry condition 
		${\bar Z}_1=-0.23+i6.53$, ${\bar Z}_2=-i 1.45$, ${\bar Z}_3=-{\bar Z}_1^*=0.23+i6.53$
		is approximately attained. (d), (e) Real and imaginary part, respectively, of the mode index, as a function of $d$,  at the operational frequency of 1.37 THz; the ${\cal PT}$-symmetric and broken phases are highlighted with different color shades.}
	\label{Figure12}
\end{figure}

\section{Possible Implementation}
\label{Sec:Implementation}
The results in Fig. \ref{Figure5} suggest some possible tunability mechanisms, which may become particularly interesting in platforms based on gate-tunable and/or photoexcited graphene sheets \cite{Bisharat:2018ml,Moccia:2020lw}. With a view toward possible implementations, it is worth recalling  that the ${\cal PT}$-symmetry condition in (\ref{eq:Zs}) imposes a rather stringent 
constraint on the gain/loss balance, which is quite challenging to attain in practice. Even from a purely theoretical viewpoint, due to causality, such condition can be exactly fulfilled only at isolated frequencies \cite{Zyablovsky:2014}. Nevertheless, some salient features of ${\cal PT}$ symmetry are still observable in imperfect conditions, with some loss offset. In fact, even {\em fully passive} configurations can be interpreted as ${\cal PT}$ -symmetric systems immersed in a dissipative background \cite{Feng:2013ed,Ornigotti:2014qp}. More broadly, EPs are generically present in many non-Hermitian systems regardless of ${\cal PT}$ symmetry, which only forces them to appear at real frequencies \cite{Miri:2019ep}. To address the practical feasibility of the required surface-impedance profiles,
we consider a typical model of a photoexcited graphene metasurface at THz frequencies \cite{Chen:2016pt}. Referring to Appendix \ref{Sec:AppB} for details, Fig. \ref{Figure12}a 
shows the schematic of such an optically pumped graphene metasurface comprising a periodic arrangement of square patches with period $D$ and separation $g$, laid on a dielectric substrate. Figures \ref{Figure12}b and \ref{Figure12}c show the normalized surface impedances of the three-section configuration in Fig. \ref{Figure1} tailored so as to approximately attain a ${\cal PT}$-symmetry condition 
(${\bar Z}_1=-0.23+i6.53$, ${\bar Z}_2=-i 1.45$, ${\bar Z}_3=-{\bar Z}_1^*=0.23+i6.53$)
at 1.37 THz. 
As shown in Figs. \ref{Figure12}d and \ref{Figure12}e, this more realistic configuration exhibits a phase transition and the occurrence of an EP, similarly to our idealized example in Fig. \ref{Figure2}. 
The required surface-impedance profile can be obtained via a spatial tailoring of the patterning together with selective doping and/or optical pumping (see Appendix \ref{Sec:AppB} for details). We stress that the values above only serve to demonstrate that the required impedance parameters are in line with those attainable from typical models in the literature, and that further studies are needed for designing a practical implementation.

\section{Conclusions and Perspectives}
\label{Sec:Conclusions}
In summary, we have theoretically demonstrated the occurrence of EPs in flat-optics scenarios, based on coupled LWs in non-Hermitian surface-impedance junctions. 
By linking the conventional \cite{Bisharat:2017ge} and non-Hermitian \cite{Moccia:2020lw} versions of LWs, supported by discontinuities in the surface reactance and resistance, respectively, our work provides a unified framework to study different 1-D waveguiding mechanisms, shedding further light on the physics of LWs, and further exploring the opportunities of this emerging research area, which is still in its infancy. Specifically, our results open up some avenues in the largely unexplored field of non-Hermitian flat optics. The inherent field enhancement and extreme localization of LWs \cite{Bisharat:2017ge,Moccia:2020lw} in conjunction with the enhanced sensitivity to perturbations occurring at an EP \cite{Wiersig:2020ro} may find important applications to chemical and biological sensing. Moreover, EPs have been successfully applied to mode discrimination in multimode laser cavities \cite{Miri:2019ep}, and may likewise find possible applications to flat-optics lasers. Possible implementations at THz frequencies based on photoexcited graphene metasurfaces seem within reach with current or near-future technologies. The dependence of the EP on the metasurface resistance and reactance profiles may allow dynamically tuning the location of the EP at will, and even define arbitrary paths in the parameter space around an EP. Of particular interest for future research is the study of longitudinal adiabatic modulations of the impedance parameters in order to encircle an EP and attain asymmetric mode switching as in Ref. \cite{Doppler:2016de}, but realized on a fully flat-optics platform, of potential interest for on-chip optical signal processing. 
More broadly, by connecting several topics of recent interest -- LWs, flat optics/electromagnetics, and non-Hermitian wave-physics -- our work may open some unexplored opportunities for extreme field localization and waveguiding on low-dimensional structures.

\appendix

\section{Details on modeling}
\label{Sec:AppA}
Unlike the single-discontinuity cases \cite{Bisharat:2017ge,Moccia:2020lw}, our geometry here (Fig. \ref{Figure1}) does not admit an analytical solution \cite{Kong:2019at}, and therefore 
the modeling in our study relies on the finite-element-based commercial software package COMSOL Multiphysics \cite{COMSOL:2015}. 
Specifically, for the eigenmode analysis (Figs. \ref{Figure2}--\ref{Figure6}, \ref{Figure8}, and \ref{Figure12}), we rely on the ``Mode Analysis'' study available in the RF Module, and assume a 2-D computational domain delimited by a semicircle of radius $2\lambda$ with a perfectly matched layer (PML) termination of thickness $0.25\lambda$ in the vacuum region and the impedance boundary condition at $y=0$. This latter is enforced by defining an electric surface current density with components $J_x = E_x/Z\left(x\right), J_y = 0, J_z = E_z/Z\left(x\right)$, with $Z\left(x\right)$ defined in (\ref{eq:Zs}).
For the domain discretization, we utilize an adaptive mesh with element size $\le0.001\lambda$ close to the impedance surface, and $\le0.03\lambda$ away from it; this
results in $\sim 1.5$ million degrees of freedom. For the system solution, we utilize the MUMPS direct solver with default parameters.

For the source-excited configurations (Figs. \ref{Figure9} and \ref{Figure10}), we utilize instead the ``Frequency domain'' study.
This time we consider a 3-D parallelepiped computational domain of size $0.75\lambda\times\lambda\times 7.5\lambda$ (along $x$, $y$ and $z$, respectively) with a PML termination of thickness $0.25\lambda$ in the vacuum region, and the same impedance boundary condition at $y=0$. 
To mimic an infinite length (along the propagation direction $z$), both ends are terminated with fictitious sections of length $2.75\lambda$, featuring a linearly tapered impedance profile that eventually matches the vacuum characteristic impedance at the PML. The computational domain is discretized with an adaptive meshing with element size $\le0.04\lambda$ close to the impedance surface, and $\le0.09\lambda$ away from it, which results in $\sim5$ million degrees of freedom. 
In the examples in Figs. \ref{Figure9}a--\ref{Figure9}c, the structure is excited by an elementary electric dipole, with characteristics specified in the figure caption. 
For the cases in Figs. \ref{Figure9}d and \ref{Figure9}e, the circularly polarized elementary sources are synthesized via the superposition of colocated $x$ and $z$-directed elementary electric dipoles with same amplitude and $\pm90^o$ phase-lag, respectively.
For the problem solution, we utilize the Pardiso direct solver and default parameters. 

For calculating the spatial spectrum in Fig. \ref{Figure11}, we consider a longer computational domain ($15\lambda$)  along the propagation direction $z$ in order to improve the resolution, and 
apply a by means of a 2048-point fast Fourier transform implemented in a Python code via the routine \texttt{fft} available in the NumPy package \cite{Harris:2020ap}.

\section{Details on Photoexcited Graphene Metasurfaces}
\label{Sec:AppB}
At terahertz frequencies, an optically pumped graphene monolayer can exhibit population inversion and hence negative dynamic conductivity. 
Following \cite{Kar:2014tp}, we model the complex conductivity as the sum of the contributions from the interband and intraband transitions
\beq
{\sigma _g} = {\sigma _{{\rm{inter}}}} + {\sigma _{{\rm{intra}}}},
\eeq
where
\begin{subequations}
	\beq
	{\sigma _{{\rm{intra}}}}\!=\!- i\frac{{{q^2}}}{{\pi \hbar }}\frac{1}{{\left( {\omega  \!+\! i{\tau ^{ - 1}}} \right)}}\int_0^\infty  {\varepsilon \!\left[ {\frac{{\partial {F_1}\left( \varepsilon  \right)}}{{\partial \varepsilon }} \!-\! \frac{{\partial {F_2}\left( { \!-\! \varepsilon } \right)}}{{\partial \varepsilon }}} \right]\!d\varepsilon } ,
	\eeq
	\beq
	{\sigma _{{\rm{inter}}}} \!=\! i\frac{{{q^2}}}{{\pi \hbar }}\left( {\omega  \!+\! i{\tau ^{ - 1}}} \right)\int_0^\infty  {\frac{{{F_2}\left( { - \varepsilon } \right) \!-\! {F_1}\left( \varepsilon  \right)}}{{{{\left( {\omega  + i{\tau ^{ - 1}}} \right)}^2} \!-\! 4{{\left( {{\varepsilon  \mathord{\left/
									{\vphantom {\varepsilon  \hbar }} \right.
									\kern-\nulldelimiterspace} \hbar }} \right)}^2}}}d\varepsilon } .
	\eeq
	\label{eq:sigmag}
\end{subequations}
In (\ref{eq:sigmag}), 
\beq
{F_1}\left( \varepsilon  \right) = \frac{1}{{1 + \exp \left( {\frac{{\varepsilon  - E_F^n}}{{{k_B}T}}} \right)}},\quad {F_2}\left( \varepsilon  \right) = \frac{1}{{1 + \exp \left( {\frac{{\varepsilon  - E_F^p}}{{{k_B}T}}} \right)}}
\eeq
are the Fermi-Dirac distributions, $q$   the electron charge, $\varepsilon$  the energy, $\hbar$   the reduced Planck’s constant, $k_B$  the Boltzmann constant, $T$  the absolute temperature, and $\tau$  the momentum relaxation time of charge carriers. Moreover, $E_F^{p,n} =  \pm {\varepsilon _F}$ are the quasi-Fermi levels resulting from the splitting of the Fermi level $E_F$ at the nonequilibirum condition originating from the pumped energy. In the absence of optical pumping, we obtain
$E_F^{p,n} = {E_F} = {\varepsilon _F}$ , and hence  $F_1=F_2$, which consistently yields the conventional expression of $\sigma_g$ under equilibrium conditions \cite{Hanson:2008dg}.
At terahertz frequencies, the real part of the (dominant) interband contribution can be approximated as \cite{Ryzhii:2007nd}
\beq
{\mathop{\rm Re}\nolimits} \left( {{\sigma_{{\rm{inter }}}}} \right) \approx \frac{{{q^2}}}{{4\hbar }}\tanh \left( {\frac{{\hbar \omega  - 2{\varepsilon _F}}}{{4{k_B}T}}} \right),
\eeq
which assumes negative values for  $2{\varepsilon _F} > \hbar \omega $. For intense optical pumping ($\varepsilon_F\gg k_BT$), the following approximation holds \cite{Chen:2016pt}
\beq
\frac{{{\varepsilon _F}}}{{{k_B}T}} \approx \sqrt {\frac{{{\pi ^2}}}{6} + \pi {{\left( {\frac{{\hbar {v_F}}}{{{k_B}T}}} \right)}^2}\frac{{{\tau _r}\alpha {I_{{\Omega _0}}}}}{{\hbar {\Omega _0}}}} ,
\eeq
where 
${v_F}$  is the Fermi velocity,  ${\tau _r}$
the recombination time,  $\alpha$  the optical absorption,  and  
${I_{{\Omega _0}}}$ the pump intensity at the pumping angular frequency  $\Omega_0$. 
This implies that the quasi-Fermi energy can be changed by varying the pump intensity  ${I_{{\Omega _0}}}$, the Fermi velocity ${v_F}$  and/or  the recombination time ${\tau _r}$. These latter, in turn,   can be tuned by varying the carrier concentration \cite{Elias:2011dc,Rana:2011uc,Kar:2014tp}. 

To increase the degree of freedom in the design, as schematically illustrated in Fig. \ref{Figure12}a , we consider a graphene metasurface comprising a periodic arrangement of square patches with period $D$ and separation $g$, laid on a dielectric substrate of relative permittivity $\varepsilon_r$. As shown in Refs. \cite{Luukkonen:2008as,Padooru:2013dc}, the surface impedance for this type class of metasurfaces can be approximately expressed as
\beq
Z\approx\frac{D}{(D-g) \sigma_g}+i \frac{\pi\eta}{2 k D \varepsilon_{eff} \log \left[\csc \left(\displaystyle{\frac{\pi g}{2 D}}\right)\right]},
\label{eq:Zms}
\eeq
with $\varepsilon_{eff}=\left(1+\varepsilon_r\right)/2$ denoting an effective relative permittivity. The curves in Figs. \ref{Figure12}b and \ref{Figure12}c are obtained from (\ref{eq:Zms}), by assuming $\varepsilon_r=3.8$ (SiO$_2$ substrate) and the photoexcited-graphene model in (\ref{eq:sigmag}), with different values of the geometrical parameters $D$ and $g$, and the quasi-Fermi energy level, so as to synthesize the three surface-impedance sections in accord with the ${\cal PT}$-symmetric distribution in (\ref{eq:Zs}). Specifically, for $Z_1$, we select 
$D=2.5\mu$m, $g=1\mu$m and $\varepsilon_F=16$meV; for $Z_2$:  
$D=6\mu$m, $g=2.5\mu$m and $\varepsilon_F=19$meV; for $Z_3$: $D=4\mu$m, $g=1.5\mu$m and $\varepsilon_F=50$meV.
The three different quasi-Fermi levels are in line with those in the literature \cite{Chen:2016pt,Elias:2011dc,Rana:2011uc,Kar:2014tp}, and
could be realized, in principle, via spatially selective doping and/or optical pumping.


%

\end{document}